\documentclass[12pt,showpacs]{revtex4}
\usepackage[english]{babel}
\usepackage{graphicx}
\usepackage{epsfig}
\begin{document}
\def\sl#1{\slash{\hspace{-0.2 truecm}#1}}
\def\beqn{\begin{eqnarray}}
\def\eeqn{\end{eqnarray}}
\def\nn{\nonumber}
%
%\draft
%

\def\eqcm{\: ,}           % punctuation in equations
\renewcommand{\d}{{\rm d}}
\newcommand{\Ta}{H_T}
\newcommand{\Tb}{\tilde{H}_T}
\newcommand{\Tc}{E_T}
\newcommand{\Td}{\tilde{E}_T}
\def\be{\begin{equation}}
\def\ee{\end{equation}}
\def\bea{\begin{eqnarray}}
\def\eea{\end{eqnarray}}
\def\ket#1{\hbox{$\vert #1\rangle$}}   % definition of ket
\def\bra#1{\hbox{$\langle #1\vert$}}   % definition of bra
\def\oneh{{\textstyle {1\over 2}}}
\def\onet{{\textstyle {1\over 3}}}
\def\smoneh{{\scriptstyle {1\over 2}}}
\def\onesix{{\textstyle {1\over 6}}}
\def\oneq{{\textstyle {1\over 4}}}
\def\treh{{\textstyle {3\over 2}}}
\def\treq{{\textstyle {3\over 4}}}
\def\oneight{{\textstyle {1\over 8}}}
\def\onesq{{\textstyle {1\over \sqrt{2}}}}

\def\Re{\hbox{\rm Re\,}}
\def\Im{\hbox{\rm Im\,}}

\def\Tr{\hbox{\rm Tr\,}}
\def\Sp{\hbox{\rm Sp\,}}
\def          % circa \ge
\simeq{{\ \lower2pt\hbox{$-$}\mkern-13mu \raise2pt \hbox{$\sim$}\ }} 
\def\dirac#1{\slash \mkern-10mu #1}

\title{Chiral-odd generalized parton distributions in constituent 
quark models
}
\author{B.~Pasquini, M.~Pincetti, S.~Boffi
}
\affiliation{Dipartimento di Fisica Nucleare e Teorica, Universit\`a degli
Studi di Pavia and INFN, Sezione di Pavia, Pavia, Italy}
\date{\today}
\begin{abstract}
{
We derive the overlap representation of chiral-odd generalized parton distributions using the Fock-state decomposition in the transverse-spin basis. This formalism is applied to the case of light-cone wave functions in a constituent quark model. Numerical results for the four chiral-odd generalized parton distributions at the hadronic scale are shown in different kinematics. In the forward limit we derive the transversity distribution, the tensor charge and the angular momentum sum rule for quarks with transverse polarization in an unpolarized nucleon.
}
\end{abstract}
\maketitle 
\section{Introduction}

The study of the hadron structure in terms of quarks and gluons, the fundamental degrees of freedom in quantum chromodynamics (QCD), still rises open and interesting questions. In high-energy processes the quark-gluon structure of the nucleon is described by a set of parton distributions. At the level of leading twist a complete quark-parton model of the nucleon requires three quark distributions, $f_1$, $g_1$, and $h_1$. The quark density, or unpolarized distribution, $f_1(x)$ is the probability of finding a quark with a fraction $x$ of the longitudinal momentum of the parent nucleon, regardless of its spin orientation. The longitudinal polarization, or helicity, distribution $g_1(x)$ measures the net helicity of a quark in a longitudinally polarized nucleon. In a transversely polarized nucleon, the transverse polarization, or transversity, distribution $h_1(x)$ is the number density of quarks with polarization parallel to that of the nucleon, minus the number density of quarks with antiparallel polarization. The first two distributions are well known quantities and can be extracted from inclusive deep-inelastic scattering (DIS) data. The last one is totally unknown because, being a chiral-odd quantity, does not contribute to inclusive DIS and is only accessible experimentally when coupled to another chiral-odd partner in the 
cross section. Several ways have been suggested to measure $h_1$~\cite{barone,varia}. These include the transversely polarized Drell-Yan process~\cite{ralston,jaffe91,ralston2,jaffe92,pax}, the single-spin asymmetry in semi-inclusive DIS~\cite{artru} and pp scattering~\cite{collins}, and the semi-inclusive reaction with two-meson interference fragmentation~\cite{tang,radici,bacchetta}.

More recently, generalized parton distributions (GPDs) have been defined~\cite{muller,radyushkin96,ji78} as non-diagonal hadronic matrix elements of bilocal products of the light-front quark and gluon field operators. They depend on the momentum transferred to the parton, as well as on the average longitudinal momentum, and contain a wealth of information about the internal structure of hadrons, interpolating between the  inclusive physics of parton distributions and the exclusive limit of electroweak form factors (for recent reviews, see e.g.~\cite{goeke,diehl03,ji04,BR05}). A complete set of quark GPDs at leading twist include four helicity conserving, usually labeled $H$, $E$, $\tilde H$, $\tilde E$, and four helicity flipping (chiral-odd) GPDs, labeled $H_T$, $E_T$, $\tilde H_T$, $\tilde E_T$~\cite{HJ98,diehl01}. In the forward limit, as the momentum transfer vanish,  $H$, $\tilde H$ and $H_T$ reduce to $f_1$, $g_1$ and $h_1$, respectively. Deeply virtual Compton scattering and hard exclusive meson production can give information about the helicity conserving GPDs, and the first experiments have been planned~\cite{exp_pl1,exp_pl2}  and/or performed~\cite{exp1,exp2}. At present there is only one proposal to give access to the chiral-odd GPDs in diffractive double meson production~\cite{ivanov}. Although it is not obvious how the chiral-odd GPDs can be directly measured in an experiment, they provide valuable information about the correlation between angular momentum and spin of quarks inside the nucleon~\cite{diehl05,Bur05}.
 
Starting from first principles as in lattice QCD one can calculate the Mellin moments of GPDs, and first results for the chiral-odd ones have been presented~\cite{gockeler, ADH97}. A variety of model calculations is available for the helicity conserving GPDs~\cite{goeke,diehl03,ji04,BR05}. Less attention has been paid up to now to the chiral-odd case. In a simple version of the MIT bag model assuming SU(6) wave functions for the valence quarks in the proton~\cite{scopetta}, only the generalized transversity distribution $H_T$ was found to be nonvanishing. 

In the present paper the chiral-odd GPDs are studied in the overlap representation of light-cone wave functions (LCWFs) that was originally proposed in Refs.~\cite{diehl2,brodsky} within the framework of light-cone quantization. In a fully covariant approach the connection between the overlap representation of GPDs and the non-diagonal one-body density matrix in momentum space has further been explored in Ref.~\cite{BPT03} making use of the correct transformation of the wave functions from the  (canonical) 
instant-form to the (light-cone) front-form description. In this way the lowest-order Fock-space components of LCWFs with three valence quarks are directly linked to wave functions derived in constituent quark models (CQMs). Results for the four helicity conserving GPDs have been 
obtained~\cite{BPT03,BPT04}, automatically fulfilling the support condition and the particle number and momentum sum rules. Important dynamical effects are introduced by the correct relativistic treatment; as a consequence, e.g., a nonzero anomalous magnetic moment of the nucleon is obtained even when all the valence quarks are accommodated in the $s$-wave. An effective angular momentum, as required by the arguments of Refs.~\cite{BHMS,burkardt05}, is introduced by the boost from the rest frame to the light-front frame producing a nonvanishing unpolarized nonsinglet (helicity-flip) quark distribution.

The paper is organized as follows. In Section 2 the relevant definitions are summarized and the derivation of the overlap representation for the chiral-odd GPDs is presented. In Section 3 we limit ourselves to discuss the valence-quark contribution obtained  with LCWFs in a CQM showing the corresponding results for the four chiral-odd GPDs. The forward limit is discussed in the next Section focusing on the transversity distribution, the tensor charge and the angular momentum sum rule for quarks with transverse polarization in an unpolarized nucleon. Concluding remarks are given in the final Section. In an Appendix we give some technical details useful for the explicit calculation with light-front CQM. 

%%%%%%%%%%%%%%%%%%%%%%%%%%%%%%%%%%%%%%%%%%%%%%

\section{Chiral-odd generalized parton distributions}

The chiral-odd GPDs are defined as non-forward matrix elements of light-like correlation functions of the 
tensor current
\begin{eqnarray}
\label{eq:gpd_odd}
& &\frac{1}{2}\int\frac{\mathit{d}z^-}{2\pi}e^{i \bar xP^+z^-}\langle
p', \lambda'|\bar{\psi}(-{z}/2)\sigma^{+i}\gamma_5
\psi({z}/2)|p, \lambda\rangle_{|_{z^+ = 0, \vec z_\perp=
0}}
\nonumber\\
& &= \frac{1}{2P^+}\bar{u}(p',
\lambda')\bigg[H^q_T\sigma^{+i}\gamma_5 +
\tilde{H}^q_T\frac{\epsilon^{+i\alpha\beta}\Delta_\alpha
P_\beta}{M^2} +
E^q_T\frac{\epsilon^{+i\alpha\beta}\Delta_\alpha\gamma_\beta}{2M}
+
\tilde{E}^q_T\frac{\epsilon^{+i\alpha\beta}P_\alpha\gamma_\beta}{M}\bigg]u(p,
\lambda),\nonumber\\
\end{eqnarray}
where $i=1,2$ is a transverse index, and $p$ $(p')$ and $\lambda$ $(\lambda')$ are the momentum and the helicity of the initial (final) proton, respectively. In the definition~(\ref{eq:gpd_odd}) we adopted the conventions of Ref.~\cite{diehl01}, i.e. the average momentum transfer is given by $P^\mu=\oneh(p+p')^\mu$, the momentum transfer is $\Delta^\mu=p'^\mu-p^\mu$, the invariant momentum square is $t=\Delta^2$ and the skewness parameter is $\xi=-\Delta^+/2P^+$. We also use the notation $v^\mu=[v^+,v^-,\vec v_\perp]$ for any four-vector $v^\mu$ with light-cone components $v^\pm=(v^0\pm v^3)/\sqrt{2}$ and $\vec v_\perp=(v^1,v^2)$. The link operator normally needed to make the definition (\ref{eq:gpd_odd}) gauge invariant does not appear because we choose the gauge $A^+=0$ and assume
that one can ignore the recently discussed transverse components of the gauge field~\cite{brodsky2002,belitsky2003}.

The chiral-odd GPDs are off-diagonal in the parton helicity basis. They become diagonal if one changes basis from eigenstates of helicity to eigenstates of transversity. As the transversity basis turns out to be rather convenient to derive the overlap representation of the chiral-odd GPDs in terms of LCWFs, it is worthwhile to show explicitly how the matrix elements which enter into the definition of the chiral-odd GPDs transform  from the helicity basis into the tranversity basis. 

According to Ref.~\cite{diehl01}, the GPDs with helicity flip can be related to the following matrix elements
\begin{eqnarray}
  \label{on-shell}
A_{\lambda'+, \lambda-} &=&
\int \frac{d z^-}{2\pi}\, e^{i\bar x P^+ z^-}
  \langle p',\lambda'|\, {\cal O}_{+,-}(z)
  \,|p,\lambda \rangle \Big|_{z^+=0,\, \vec{z}_\perp=0} \, ,
\nonumber \\
A_{\lambda'-, \lambda+} &=&
\int \frac{d z^-}{2\pi}\, e^{i\bar x P^+ z^-}
  \langle p',\lambda'|\, {\cal O}_{-,+}(z)
  \,|p,\lambda \rangle \Big|_{z^+=0,\, \vec{z}_\perp=0} \, ,
\nonumber \\
\end{eqnarray}
with the operators $O_{+,-}$ and $O_{-,+}$ defined by
\begin{eqnarray}
{\cal O}_{+,-} 
&=& \frac{i}{4}\, 
  \bar{\psi}\, \sigma^{+1} (1-\gamma_5)\, \psi \,,\nonumber\\
{\cal O}_{-,+} 
&=& - \frac{i}{4}\, \bar{\psi}\, \sigma^{+1} (1+\gamma_5)\, \psi.
\hspace{2em}
\label{eq:op_o}
\end{eqnarray}

By using the definitions in Eqs.~(\ref{eq:gpd_odd}) and (\ref{on-shell}) and working in the reference frame where  the momenta $\vec p$ and $\vec p\,'$ lie in the $x-z$ plane, one can derive the following relations~\cite{diehl01}
\begin{eqnarray}
  \label{flip-amplitudes}
A_{++,+-} &=&   \epsilon\, \frac{\sqrt{t_0-t}}{2m} \left( \Tb^q
              + (1-\xi)\, \frac{\Tc^q + \Td^q}{2} \right) , 
\nonumber \\
A_{-+,--} &=&   \epsilon\, \frac{\sqrt{t_0-t}}{2m} \left( \Tb^q
              + (1+\xi)\, \frac{\Tc^q - \Td^q}{2} \right) , 
\nonumber \\
A_{++,--} &=& \sqrt{1-\xi^2} \left(\Ta^q + \
              \frac{t_0-t}{4 m^2}\, \Tb^q -
              \frac{\xi^2}{1-\xi^2}\, \Tc^q +
              \frac{\xi}{1-\xi^2}\, \Td^q \right) , 
\nonumber \\
A_{-+,+-} &=& - \sqrt{1-\xi^2}\; \frac{t_0-t}{4 m^2}\, \Tb^q,
\end{eqnarray}
where one has used the relation 
$A_{-\lambda'-,-\lambda +}=(-1)^{\lambda'-\lambda}A_{\lambda'+,\lambda -}$ 
due to parity invariance. In Eqs.~(\ref{flip-amplitudes}), 
$- t_0 = 4 m^2 \xi^2/(1-\xi^2)$ 
is the minimum value of $-t$ for given $\xi$, and $\epsilon =
\mathrm{sgn}(D^1)$, where $D^1$ is the $x$-component of $D^\alpha =
P^+ \Delta^\alpha - \Delta^+ P^\alpha$.

In the framework of light-cone quantization, the independent dynamical fields are the so-called ``good'' LC components of the fields, namely $\phi=P_+\psi$ with the projector $P_+=\oneh \gamma^-\gamma^+$. By introducing the helicity basis given by the right-handed ($R$) and left-handed ($L$) projections of the field $\phi$, namely
$\phi_R=P_R\phi=\oneh(1+\gamma_5)\phi$ and 
$\phi_L=P_L\phi=\oneh(1-\gamma_5)\phi,$ 
it is easy to see that 
\begin{equation}
O_{+,-}=\frac{1}{\sqrt{2}}\phi^\dagger_R\phi_L,\qquad
O_{-,+}=\frac{1}{\sqrt{2}}\phi^\dagger_L\phi_R.
\label{o_helicity}
\end{equation}
This last equation explicitly shows the chirally odd nature of the distributions $\Ta^q$, $\Tb^q$, $\Tc^q$, $\Td^q$. 

Alternatively, one can work in the transversity basis given by the eigenstates of the transverse-$x$ spin-projection operators,
$\mathcal{Q}_\pm=\oneh(1\pm \gamma^1\gamma_5)$~\cite{Jaffe}, 
\begin{eqnarray}
\mathcal{Q}_+\phi&\equiv&\phi_{\uparrow},
\\
\mathcal{Q}_-\phi&\equiv&\phi_{\downarrow},
 \end{eqnarray}
where $\uparrow$ ($\downarrow$) is directed along (opposite to) the transverse direction $\hat x$.
In this basis, it is convenient to consider the following operators
\begin{eqnarray}
\mathcal{N}_T&=&
\mathcal{O}_{+,-} + \mathcal{O}_{-,+}=
-\frac{i}{2}\bar{\psi}\sigma^{+1}\gamma_5\psi
=\frac{1}{\sqrt{2}}\phi^\dagger\gamma^1\gamma_5 \phi=
\frac{1}{\sqrt{2}}
\left(\phi^\dagger_\uparrow\phi_\uparrow
-\phi^\dagger_\downarrow\phi_\downarrow\right),\label{eq:nf}\\
\mathcal{F}_T&=&
\mathcal{O}_{+,-} - \mathcal{O}_{-,+} = \frac{i}{2}\bar{\psi}\sigma^{+1}\psi
=-\frac{1}{\sqrt{2}}\phi^\dagger\gamma^1 \phi
= \frac{1}{\sqrt{2}}\left(
\phi^\dagger_\downarrow \phi_\uparrow
-\phi^\dagger_\uparrow \phi_\downarrow \right).\label{eq:flip}
\end{eqnarray}
We note that  the operator $\mathcal{N}_T$ is given by difference of the density operators for $\uparrow$  and $\downarrow$ projections of the transverse polarization, while the off-diagonal matrix elements 
of  the density matrix of the spin in the transverse $\hat x$ direction  appear in the operator $\mathcal{F}_T$
\footnote{Analogously, by working in the basis of eigenstates of the spin-projection operators in the transverse $\hat y$ direction, one finds that $\mathcal{N}_T$ is related to the off-diagonal matrix elements of the spin matrix in the transverse $\hat y$ direction, while $\mathcal{F}_T$ is given in terms of the density operators for polarization in the positive and negative $\hat y$
direction.}. We now introduce the transversity basis for the nucleon spin states, i.e.
\begin{eqnarray} 
|p,\uparrow\rangle=\frac{1}{\sqrt{2}}(|p,+\rangle+|p,-\rangle)
\nonumber\\
|p,\downarrow\rangle=\frac{1}{\sqrt{2}}(|p,+\rangle -|p,-\rangle),
\label{eq:trans_spinor}
\end{eqnarray}
and define the following matrix elements
\begin{eqnarray}
\label{eq:T}
T^{q}_{\lambda'_t\lambda_t} &=& \langle p',
\lambda'_t|\int\frac{dz^-}{2\pi}e^{i \bar xP^+z^-}
\bar{\psi}(-{z}/2)\gamma^+\gamma^1\gamma_5
\psi({z}/2)|p, \lambda_t\rangle,
\\
\label{eq:Ttilde}
\tilde{T}^{q}_{\lambda'_t\lambda_t} &=&\langle p',
\lambda'_t|\int\frac{dz^-}{2\pi}e^{i\bar xP^+z^-}
\frac{i}{2}\bar{\psi}(-{z}/2)\sigma^{+1}\psi({z}/2)
|p, \lambda_t\rangle,
\end{eqnarray}
where $\lambda_t$ ($\lambda'_{t}$) labels the transverse polarization of the initial (final) nucleon in the $\uparrow$ or $\downarrow$ direction. Due to parity invariance these matrix elements obey the following relations
\begin{eqnarray}
T^{q}_{\uparrow\uparrow}=-T^{q}_{\downarrow\downarrow}\,,&\quad&
T^{q}_{\uparrow\downarrow}=T^{q}_{\downarrow\uparrow}\,,\nonumber\\
\tilde T^{q}_{\uparrow\uparrow}=\tilde T^{q}_{\downarrow\downarrow}\, ,&\quad&
\tilde T^{q}_{\uparrow\downarrow}=-\tilde T^{q}_{\downarrow\uparrow}\, ,
\end{eqnarray}
and are related to the matrix elements in the helicity basis by
\begin{eqnarray}
T^q_{\uparrow\uparrow} =A_{++,--} + A_{-+,+-}, &\quad&
T^q_{\uparrow\downarrow} =A_{++,+-} - A_{-+,--},\nonumber\\
\tilde{T}^q_{\uparrow\uparrow} = 
A_{++,+-}+A_{-+,--}, &\quad&
\tilde{T}^q_{\downarrow\uparrow} = 
A_{++,--}-A_{-+,+-}.
\end{eqnarray}

Finally, the chiral-odd GPDs are obtained from the transverse matrix elements through the relations
\begin{eqnarray}
H^{q}_T &=& \frac{1}{\sqrt{1 - \xi^2}}T^q_{\uparrow\uparrow} -
\frac{2M\xi} {\epsilon\sqrt{t_0 - t}(1 - \xi^2)}T^q_{\uparrow\downarrow},
\nonumber\\
E_T^{q} &=& 
\frac{2M\xi}{\epsilon\sqrt{t_0 - t}}\frac{1}{1 - \xi^2}T_{\uparrow\downarrow}^q
+ \frac{2M}{\epsilon\sqrt{t_0 - t}(1 - \xi^2)}\tilde{T}^q_{\uparrow\uparrow}
\nonumber\\
&&- \frac{4M^2}{(t_0 - t)\sqrt{1 - \xi^2}(1 - \xi^2)}
\bigg(\tilde{T}^q_{\downarrow\uparrow} - T^q_{\uparrow\uparrow}\bigg),
\nonumber\\
\tilde{H}_T^{q} &=& \frac{2M^2}{(t_0 - t)\sqrt{1 -\xi^2}}
(\tilde{T}^q_{\downarrow\uparrow} - T^q_{\uparrow\uparrow}),\nonumber\\
\tilde{E}^{q}_T &=& \frac{2M}{\epsilon\sqrt{t_0 - t}(1 -\xi^2)}
\bigg(T^q_{\uparrow\downarrow}
+ \xi\tilde{T}^q_{\uparrow\uparrow}\bigg)\nonumber\\ &&
- \frac{4M^2\xi}{(t_0 - t)\sqrt{1 - \xi^2}(1 - \xi^2)}
\bigg(\tilde{T}^q_{\downarrow\uparrow} - T^q_{\uparrow\uparrow}\bigg).
\nonumber\\
\label{eq:gpd_oddampl}
\end{eqnarray}

\subsection{The overlap representation}

In the following we will restrict our discussion to the region $\xi \leq \bar x\leq 1$ of plus-momentum fractions, where the generalized quark distributions describe the emission of a quark with plus-momentum $(\bar x+\xi)P^+$ and its reabsorption with plus-momentum $(\bar x-\xi)P^+$.

The derivation of the overlap representation of the $T_{\lambda_t\lambda'_t}$ and  $\tilde T_{\lambda_t\lambda'_t}$ matrix elements goes along the line described in Ref.~\cite{diehl2} for the case of the matrix elements defining the GPDs in the chiral-even sector. Here we report the essential steps of the 
derivation.

At the light-cone time $z^+=0$, the Fourier components of $\phi_\uparrow$ are the annihilation operator for an on-shell quark with transverse polarization $\uparrow$ ($b_\uparrow$) and the creation operator
for an on-shell antiquark with transverse polarization $\downarrow$ ($d^\dagger_\downarrow$), i.e.
\begin{eqnarray}
\label{eq:quarkexpansion}
\phi_\uparrow(z^-,{\vec z}_\perp)&=&
\int\frac{\d k^+\,\d^2{\vec k}_{\perp}}{k^+\,16\pi^3}\,\Theta(k^+)\;
\nonumber\\ 
&& \bigg\{ 
     b_\uparrow(k^+,\vec k_\perp)\;u_+(k,\uparrow)
  \;\exp\left({}-i\,k^+z^-+i\,{\vec k}_{\perp}
    \cdot{\vec z}_\perp)\right)
\nn\\
&& \phantom{\bigg\{ }
{{} + {}} d_\uparrow^{\,\dagger}(k^+,\vec k_\perp)\,v_+(k,\uparrow)
  \;\exp\left({}+i\,k^+z^--i\,
    {\vec k}_{\perp}\cdot{\vec z}_\perp)\right)
\bigg\} ,
\end{eqnarray} 
where $u_+(k,\uparrow)=P_+u(k,\uparrow)$ and $v_+(k,\uparrow)=P_+ v(k,\uparrow)$ are the projections into the ``good'' components of the quark and antiquark spinor. The Fourier decomposition of the field $\phi_\downarrow$ is simply obtained from Eq.~(\ref{eq:quarkexpansion}) with the replacement 
$\uparrow\ \leftrightarrow\ \downarrow$. The Fock-space in the transversity basis can be constructed
by successive applications to the vacuum state of the $d^\dagger_{\uparrow(\downarrow)}$ and 
$b^\dagger_{\uparrow(\downarrow)}$ operators. In this space, the representation of the nucleon state 
reads
\begin{equation}
\label{eq:Fockstate}
\left|p,\lambda_t\right\rangle = \sum_{N,\beta} 
\int [\d x]_N [\d^2 {\vec k}_\perp]_N\;
\Psi_{\lambda_t ,N,\beta}(r) \;
\left|N,\beta;k_1,\ldots,k_N \right\rangle \eqcm
\label{eq:representation}
\end{equation} 
where $\Psi_{\lambda_t ,N,\beta}$ 
is the momentum LCWF of the $N$-parton Fock state $|N,\beta;k_1,\ldots,k_N \rangle$.  The integration measures in Eq.~(\ref{eq:representation}) are defined as
\be
[\d x]_N = \prod_{i=1}^N dx_i \,\delta\left(1-\sum_{i=1}^N x_i\right), \qquad
 [\d^2 {\vec k}_\perp]_N = \frac{1}{(16\pi^3)^{N-1}}\prod_{i=1}^N\delta^2\left(\sum_{i=1}^N{\vec k}_{\perp,i} - {\vec p}_\perp\right).
 \ee
The argument $r$ of the LC wave function represents the set of kinematical variables of the $N$ partons, while  the index $\beta$ labels the parton composition and the transverse spin of each parton. Finally, replacing Eqs.~(\ref{eq:quarkexpansion}) and (\ref{eq:Fockstate}) in the expressions ~(\ref{eq:T}) and (\ref{eq:Ttilde}) for the matrix elements $T_{\lambda_t\lambda'_t}$ and 
$\tilde T_{\lambda_t\lambda'_t},$
one finds
\begin{eqnarray}
\label{eq:T_ov}
T^{q }_{\lambda'_t\lambda_t} &=&
\sum_{N,\beta=\beta'}
\bigg(\sqrt{1 - \xi}\bigg)^{2 - N}\bigg(\sqrt{1 + \xi}\bigg)^{2 -
N}
\sum_{j=1}^N \mathrm{sign}(\mu_{j}^{t})
\delta_{s_jq}\nonumber\\
&&\times\int[d\bar{x}]_N[d^2\vec{k}_\perp]_N\delta(\bar{x} -
\bar{x}_j)\mathit{\Psi}^{*}_{\lambda'_t,N,\beta'}(\hat{r}')
\mathit{\Psi}_{\lambda_t,N,\beta}(\tilde{r}),
\\
& &\nonumber\\
\label{eq:Ttilde_ov}
\tilde{T}^{q }_{\lambda'_t\lambda_t} &=&
\sum_{\beta,\beta',N}
\bigg(\sqrt{1 - \xi}\bigg)^{2 - N}\bigg(\sqrt{1 + \xi}\bigg)^{2 -
N}
\sum_{j=1}^N\delta_{{\mu_j^t}'-\mu_j^t}\delta_{{\mu_i^t}'\mu_i^t}\mathrm{sign}(\mu_{j}^{t})
\delta_{s_jq}\nonumber\\
&&\times\int[d\bar{x}]_N[d^2\vec{k}_\perp]_N\delta(\bar{x} -
\bar{x}_j)\mathit{\Psi}^{*}_{\lambda'_t,N,\beta'}(\hat{r}')
\mathit{\Psi}_{\lambda_t,N,\beta}(\tilde{r}),
\end{eqnarray}
where $s_j$ labels the quantum numbers of the $j$th active parton, with transverse initial (final) spin polarization $\mu^{t}_{j}$ (${\mu_j^t}'$),  and $\mu^{t}_{i} \, ({\mu_i^t}')$ are the transverse spin of the spectator initial (final) quarks. The set of kinematical variables $r,r'$ are defined as follows: for the final struck quark,
\be
\label{eq:finalstruck}
y'_j %= \frac{\overline k^+_j + \oneh\Delta^+}{\overline P^+ + \oneh\Delta^+}
= \frac{\displaystyle \overline x_j-\xi}{\displaystyle 1-\xi},\qquad
\vec{\kappa}'_{\perp j} = \vec{k}_{\perp j} + \oneh
\frac{\displaystyle 1-\overline x_j}{\displaystyle 1-\xi}\vec{\Delta}_\perp, 
\ee
for the final $N-1$ spectators ($i\ne j$),
\be
\label{eq:finalspect}
y'_i = \frac{\displaystyle \overline x_i}{\displaystyle 1-\xi},\qquad\vec{\kappa}'_{\perp i}
= \vec{k}_{\perp i} - \oneh
\frac{\displaystyle \overline x_i}{\displaystyle 1-\xi}\vec{\Delta}_\perp,
\ee
and for the initial struck quark
\be
\label{eq:initialstruck}
y_j %= \frac{\overline k^+_j - \oneh\Delta^+}{\overline P^+ - \oneh\Delta^+}
= \frac{\displaystyle \overline x_j+\xi}{\displaystyle 1+\xi}, 
\qquad
\vec{\kappa}_{\perp j}= \vec{k}_{\perp j} - \oneh
\frac{\displaystyle 1-\overline x_j}{\displaystyle 1+\xi}\vec{\Delta}_\perp,
\ee
for the initial $N-1$ spectators ($i\ne j$),
\be
\label{eq:initialspect}
y_i = \frac{\displaystyle \overline x_i}{\displaystyle 1+\xi},
\qquad 
\vec{\kappa}_{\perp i}=\vec{k}_{\perp i} + \oneh
\frac{\displaystyle \overline x_i}{\displaystyle 1+\xi}\vec{\Delta}_\perp.
\ee

\section{The valence-quark contribution}

In this section we specialize the results for the chiral-odd GPDs obtained above to the case of $N=3,$ 
which corresponds to truncate the Fock expansion of the nucleon state to the parton configuration given by three-valence quarks. In this framework, Eqs.~(\ref{eq:T_ov}) and (\ref{eq:Ttilde_ov})  
become
\begin{eqnarray}
\label{eq:T3_ov}
T^{q }_{\lambda'_t\lambda_t} &=&
\frac{1}{\sqrt{1 - \xi^2}}
\sum_{\mu_i^t\tau_i}\sum_{j=1}^3
\mathrm{sign}(\mu_{j}^{t})
\delta_{\tau_j\tau_q}
\int[d\overline x]_3[d\vec{k}_\perp]_3\,\delta(\overline x-\overline x_j)
\nonumber\\
& & \quad\times 
\Psi^{[f]\,*}_{\lambda'_t}(r',\{\mu_i^t\},\{\tau_i\}) 
\Psi^{[f]}_{\lambda_t}(r,\{\mu_i^t\},\{\tau_i\})\Theta(\overline x_j),
\\ 
& &\nonumber\\
\label{eq:Ttilde3_ov}
\tilde{T}^{q }_{\lambda'_t\lambda_t} &=&
\frac{1}{\sqrt{1 - \xi^2}}
\sum_{\mu_i^t\mu'_i{^t}\tau_i}\sum_{j=1}^3
\mathrm{sign}(\mu_{j}^{t})
\delta_{\tau_j\tau_q}
\delta_{{\mu_j^t}'-\mu_j^t}\delta_{{\mu_i^t}'\mu_i^t}
\int[d\overline x]_3[d\vec{k}_\perp]_3\,\delta(\overline x-\overline x_j)
\nonumber\\
& & \quad\times 
\Psi^{[f]\,*}_{\lambda'_t}(r',\{\mu^{t}_{i}\},\{\tau_i\}) 
\Psi^{[f]}_{\lambda_t}(r,\{{\mu_i^t}'\},\{\tau_i\})\Theta(\overline x_j),
\end{eqnarray}
where $\Psi^{[f]}_{\lambda_t}(r,\{\mu^{t}_{i}\},\{\tau_i\})$ is the eigenfunction of the light-front Hamiltonian of the nucleon, described as a system of three interacting quarks. It is here obtained from the corresponding solution $\Psi^{[c]}_{\lambda_t} (\{\vec{\kappa}_i\},\{\mu^{t}_{i}\},\{\tau_i\})$ of the eigenvalue equation in the instant-form as described in Ref.~\cite{BPT03}. Separating the spin-isospin component from the space part of the wave function, 
\be
\label{eq:separated}
\Psi^{[c]}_{\lambda_t} (\{\vec{\kappa}_i\},\{\mu^{t}_{i}\},\{\tau_i\})
= \psi(\vec{\kappa}_1,\vec{\kappa}_2,\vec{\kappa}_3)
\Phi_{\lambda_t\,\tau}(\mu^{t}_{1},\mu^{t}_{2},\mu^{t}_{3},\tau_1,\tau_2,\tau_3) ,
\ee
we have
\bea
\label{eq:transform}
& & \Psi^{[f]}_{\lambda_t} (r,\{\mu^{t}_{i}\},\{\tau_i\})
=  2(2\pi)^3\left[\frac{1}{M_0}\frac{\omega_1\omega_2\omega_3}
{y_1y_2y_3}\right]^{1/2} 
\psi(\vec{\kappa}_1,\vec{\kappa}_2,\vec{\kappa}_3) 
\nonumber\\
& & \qquad \times\sum_{\mu_1\mu_2\mu_3}
{D}^{1/2\,*}_{\mu^{t}_{1}\lambda^{t}_{1}}(R_{cf}(\kappa_1))
{D}^{1/2\,*}_{\mu^{t}_{2}\lambda^{t}_{2}}(R_{cf}(\kappa_2))
{D}^{1/2\,*}_{\mu^{t}_{3}\lambda^{t}_{3}}(R_{cf}(\kappa_3))
\nonumber\\
& & \qquad\ {}\times
\Phi_{\lambda_t\,\tau}(\mu^{t}_1,\mu^{t}_2,\mu^{t}_3,\tau_1,\tau_2,\tau_3) ,
\\ \nonumber
\eea
where $M_0$ is the mass of the non-interacting 3-quark system, $\omega_i=(\kappa^+_i+\kappa^-_i)/\sqrt{2}$, and the matrix ${D}^{1/2}_{\lambda_t\mu_t}(R_{cf}(k))$  are given by the representation of the Melosh rotation $R_{cf}$ in the transverse-spin space
\bea
{D}^{1/2}_{\lambda_t\mu_t}(R_{cf}(k)) & = &
\bra{\lambda_t}R_{cf}(\overline x M_0,\vec{k}_\perp)\ket{\mu_t} \nonumber\\
& = & \bra{\lambda_t}
\frac{m+\overline xM_0-i\vec{\sigma}\cdot(\hat{\vec{z}}\times\vec{k}_\perp)}
{\sqrt{(m+\overline xM_0)^2+\vec{k}_\perp^2}}\ket{\mu_t}.
\\ \nonumber
\eea
By assuming a SU(6) symmetric model for the spin-isospin component of the wave function, the summation over the spin and isospin variables in Eqs.~(\ref{eq:T3_ov}) and (\ref{eq:Ttilde3_ov}) can be cast into a rather compact analytical expression. The final results with some technical details for the derivation are reported in Appendix~\ref{appendix}.

\subsection{Results}

As an application of the general formalism developed in the previous sections we consider the valence-quark contribution to the chiral-odd GPDs calculated starting from an instant-form wave function of the proton derived in the relativistic hypercentral quark model of Ref.~\cite{FTV99}. This CQM is able to reproduce the basic features of the low-lying nucleon spectrum and was already adopted in previous studies on the helicity conserving GPDs~\cite{BPT03,BPT04}. The structure of the nucleon wave function in this model is SU(6) symmetric for the spin-isospin components and is given by Eq.~(\ref{eq:separated}). Therefore we can use the analytical expressions reported in Appendix A for the summation over spin and isospin variables, whereas the integrations over momenta are performed numerically.

%%%%%%%%%%%%%%%%%%%%%%%%%%%%%  fig. 1  %%%%%%%%%%%%%%%%%%%%%%%%%%%%%%%%%%%%%%%
\begin{figure}[ht]
\begin{center}
\epsfig{file=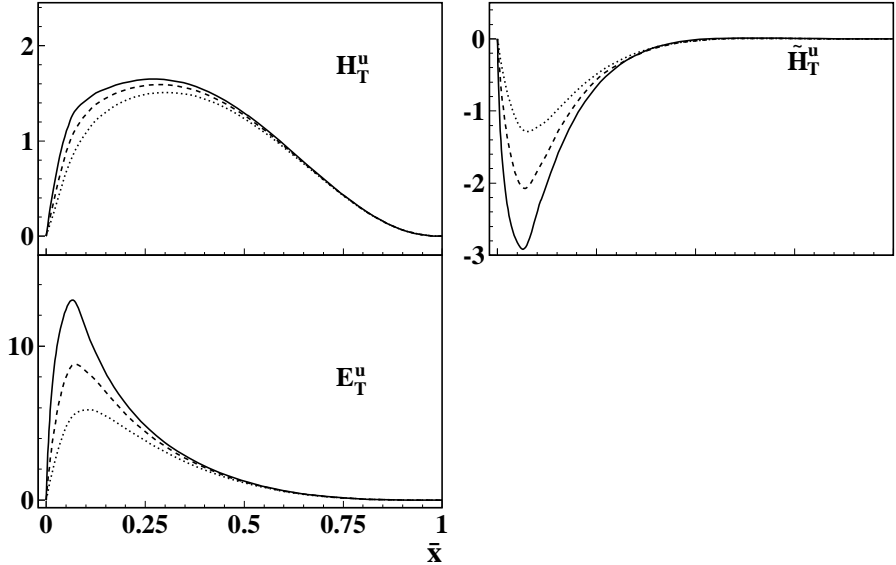,width=12 cm}
\end{center}
\vspace{-0.4cm}
\caption{\small The chiral odd generalized parton distributions calculated in the hypercentral CQM for the flavour $u$, at $\xi=0$ and different values of $t$: $t=0$ (solid curves), $t=-0.2$ (GeV)$^2$ (dashed curves), $t=-0.5$ (GeV)$^2$ (dotted curves).} 
\label{fig:fig1}
\end{figure}

%%%%%%%%%%%%%%%%%%%%%%%%%%%%%  fig. 2  %%%%%%%%%%%%%%%%%%%%%%%%%%%%%%%%%%%%%%%
\begin{figure}[ht]
\begin{center}
\epsfig{file=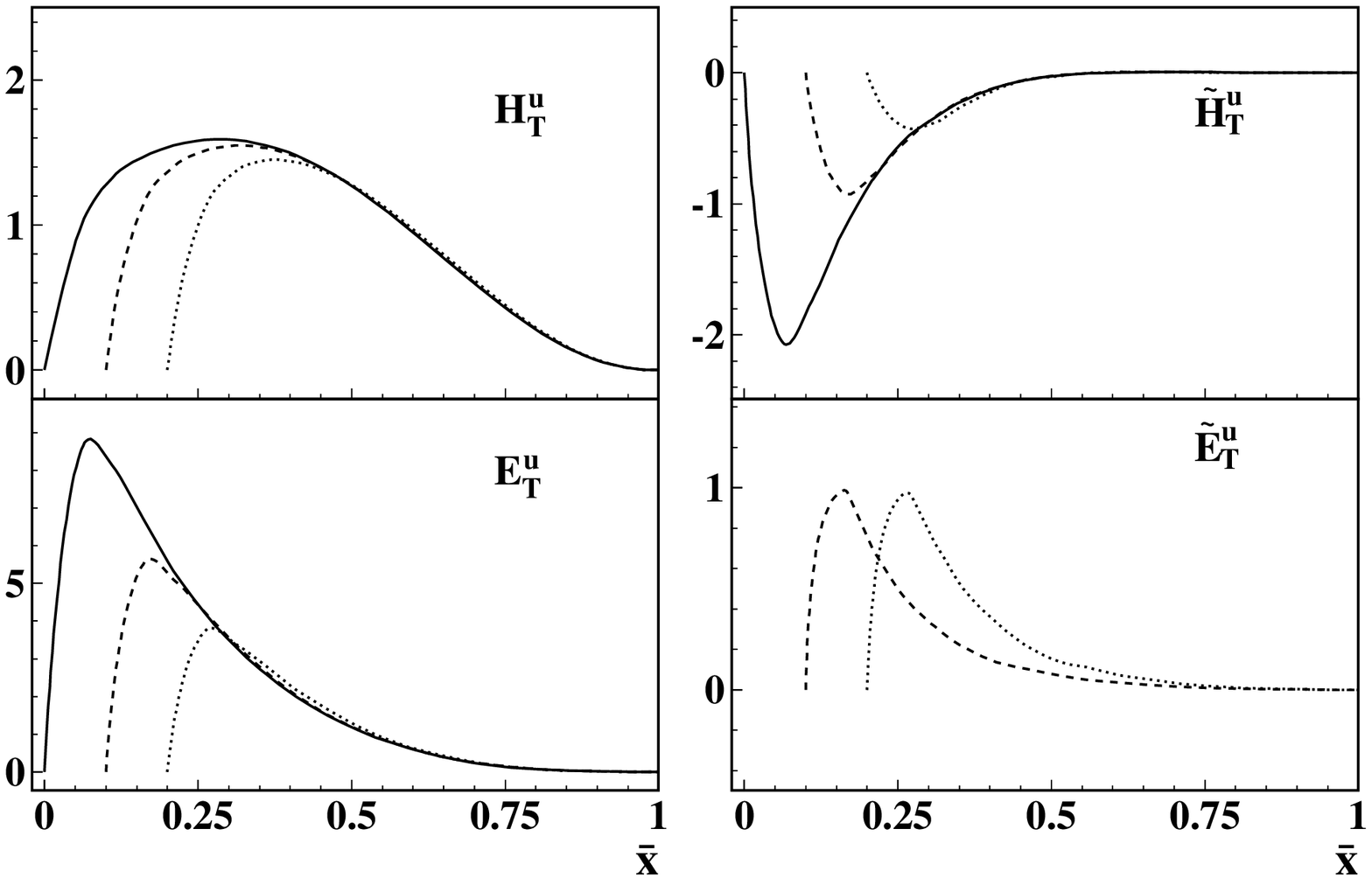,  width=26 pc}
\end{center}
\vspace{-0.4cm}
\caption{\small The same as in Fig.~\ref{fig:fig1} but for fixed $t=-0.2$ (GeV)$^2$ and different values of $\xi$: $\xi =0$ (solid curves), $\xi=0.1$ (dashed curves), $\xi=0.2$ (dotted curves).}
\label{fig:fig2}
\end{figure}

%%%%%%%%%%%%%%%%%%%%%%%%%%%%%  fig. 3  %%%%%%%%%%%%%%%%%%%%%%%%%%%%%%%%%%%%%%%
\begin{figure}[ht]
\begin{center}
\epsfig{file=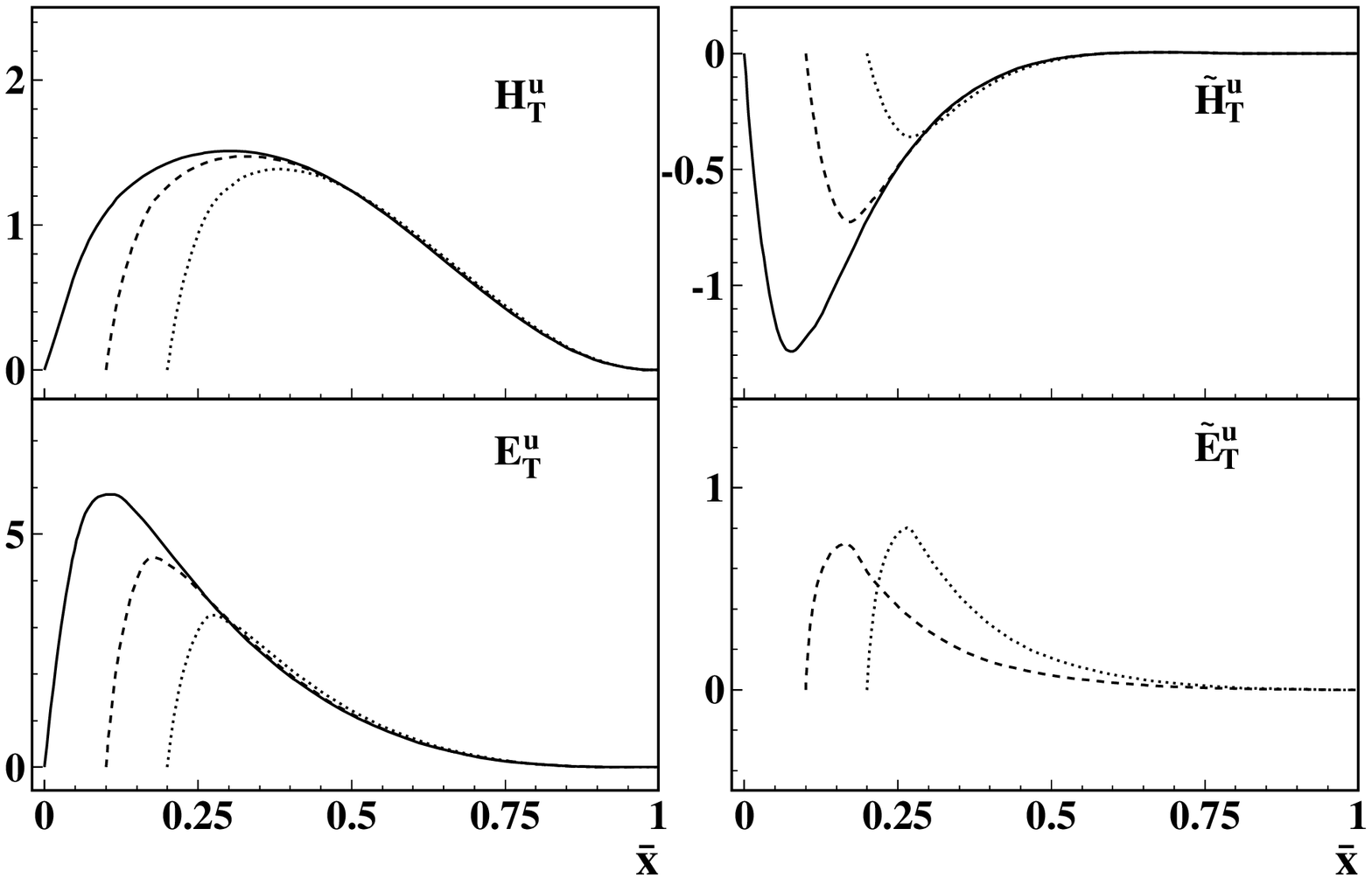,  width=26 pc}
\end{center}
\vspace{-0.4cm}
\caption{\small The same as in Fig.~\ref{fig:fig1} but for fixed $t=-0.5$ (GeV)$^2$ and different values of $\xi$: $\xi =0$ (solid curves), $\xi=0.1$ (dashed curves), $\xi=0.2$ (dotted curves).}
\label{fig:fig3}
\end{figure}

%%%%%%%%%%%%%%%%%%%%%%%%%%%%%  fig. 4  %%%%%%%%%%%%%%%%%%%%%%%%%%%%%%%%%%%%%%%
\begin{figure}[ht]
\begin{center}
\epsfig{file=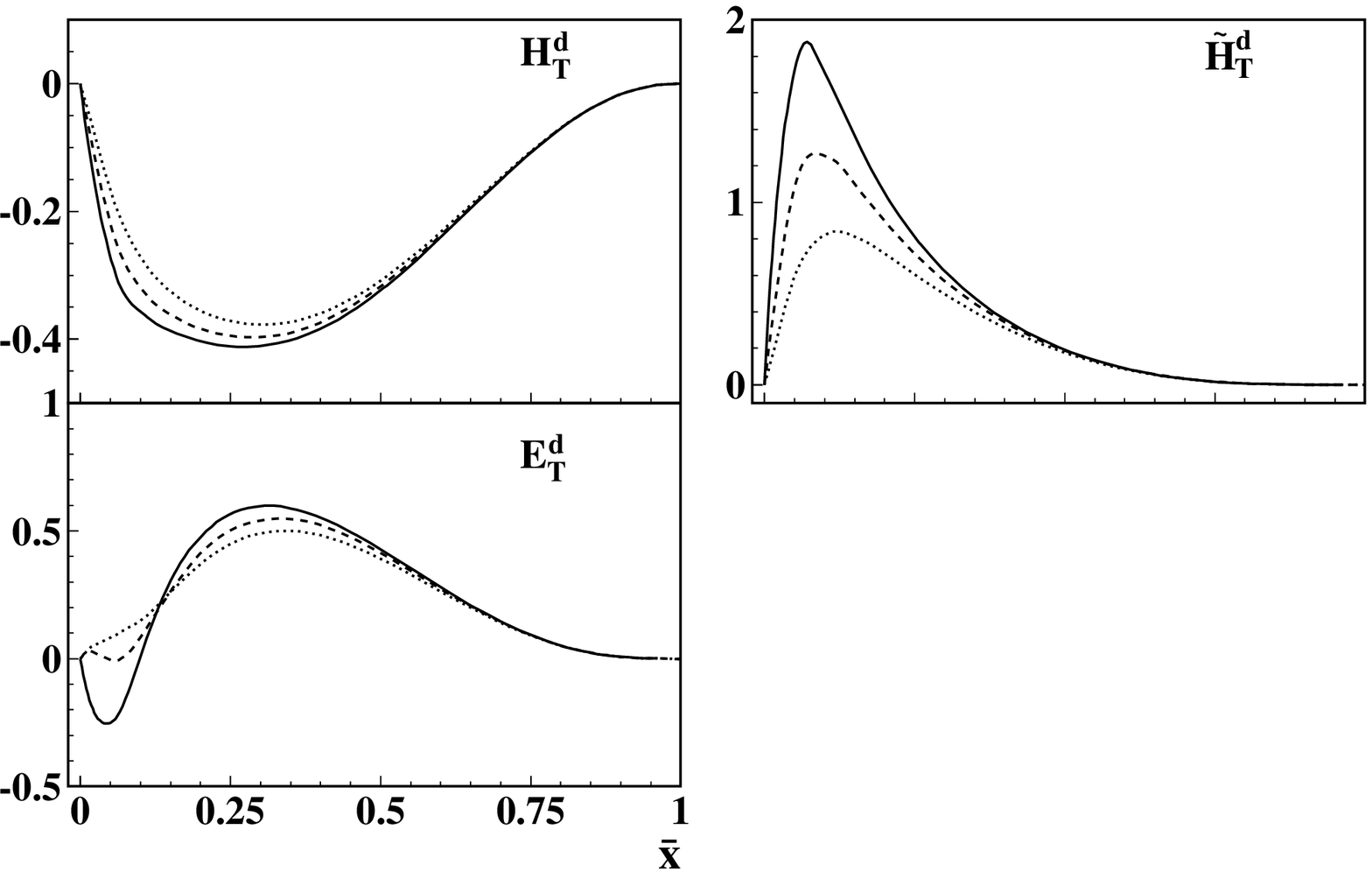,  width=26 pc}
\end{center}
\vspace{-0.4cm}
\caption{\small The chiral odd generalized parton distributions calculated in the hypercentral CQM for the flavour $d$, at $\xi=0$ and different values of $t$: $t=0$ (solid curves), $t=-0.2$ (GeV)$^2$ (dashed curves), $t=-0.5$ (GeV)$^2$ (dotted curves).} 
\label{fig:fig4}
\end{figure}
%%%%%%%%%%%%%%%%%%%%%%%%%%%%%  fig. 5 %%%%%%%%%%%%%%%%%%%%%%%%%%%%%%%%%%%%%%%
\begin{figure}[ht]
\begin{center}
\epsfig{file=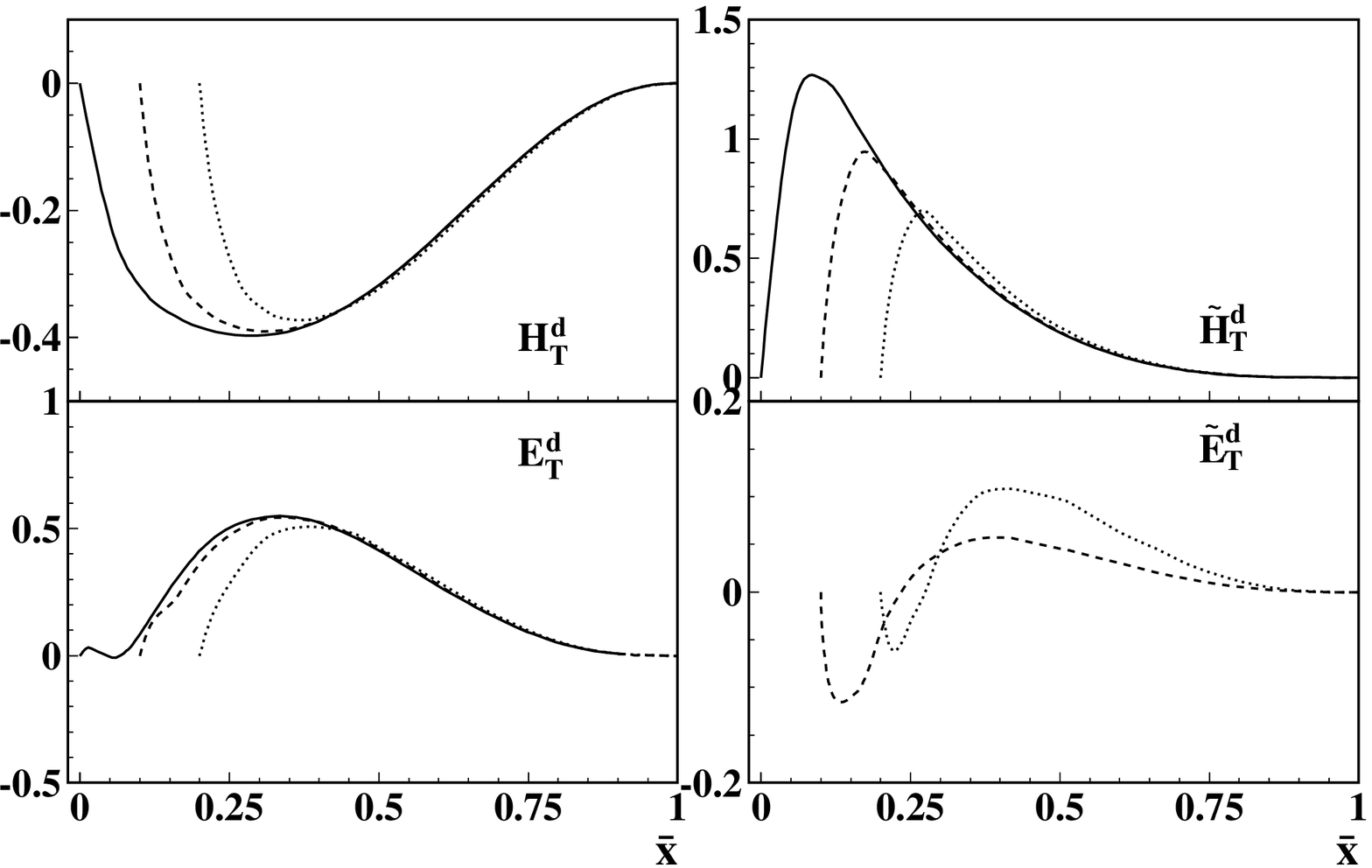,  width=26 pc}
\end{center}
\vspace{-0.4cm}
\caption{\small The same as in Fig.~\ref{fig:fig4} but for fixed $t=-0.2$ (GeV)$^2$ and different values of $\xi$: $\xi =0$ (solid curves), $\xi=0.1$ (dashed curves), $\xi=0.2$ (dotted curves).}
\label{fig:fig5}
\end{figure}

%%%%%%%%%%%%%%%%%%%%%%%%%%%%%  fig. 6  %%%%%%%%%%%%%%%%%%%%%%%%%%%%%%%%%%%%%%%
\begin{figure}[ht]
\begin{center}
\epsfig{file=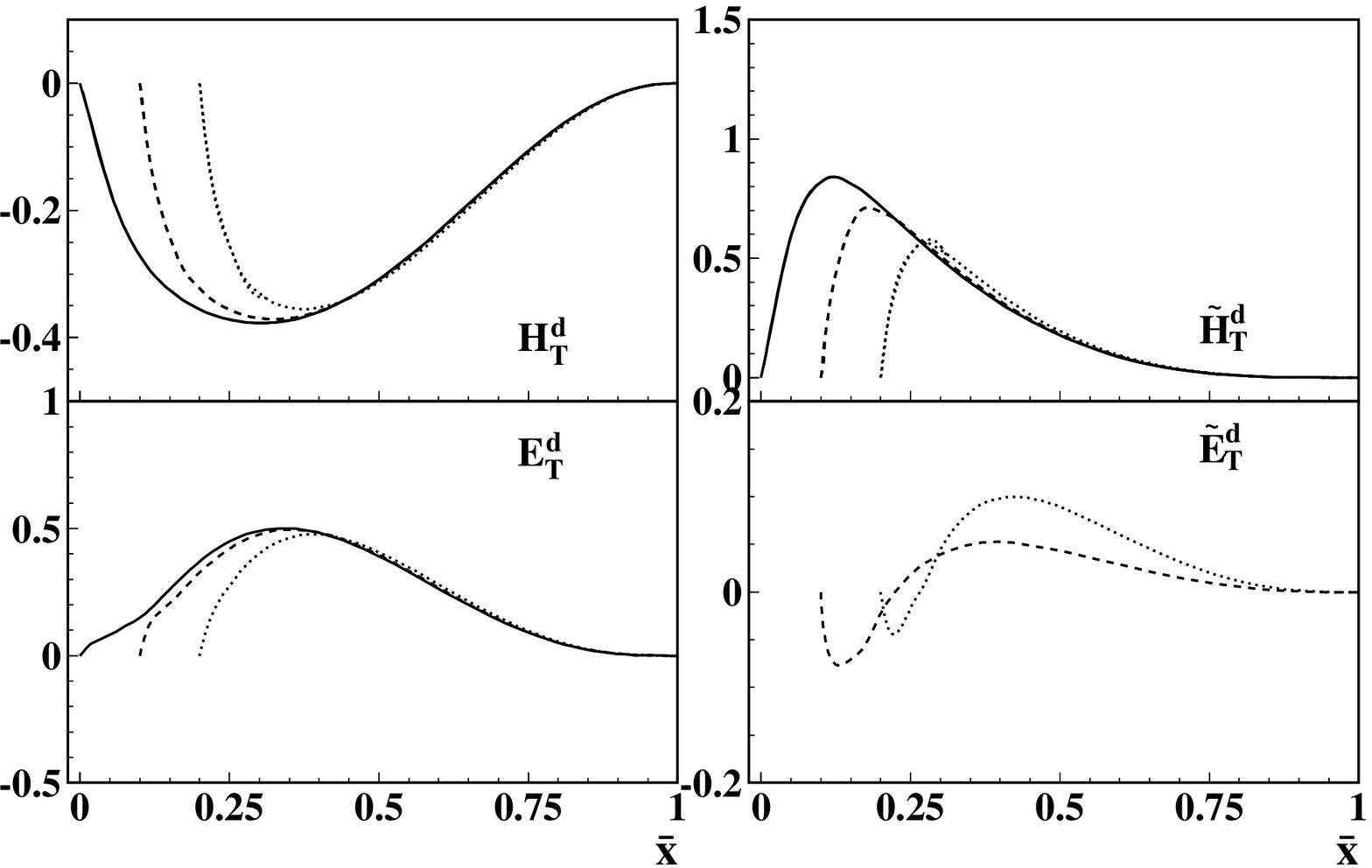,  width=26 pc}
\end{center}
\vspace{-0.4cm}
\caption{\small The same as in Fig.~\ref{fig:fig4} but for fixed $t=-0.5$ (GeV)$^2$ and different values of $\xi$: $\xi =0$ (solid curves), $\xi=0.1$ (dashed curves), $\xi=0.2$ (dotted curves).}
\label{fig:fig6}
\end{figure}

The four calculated chiral-odd GPDs, $H^q_T$, $E^q_T$, $\tilde H^q_T$, $\tilde E^q_T$, are shown in Figs.~\ref{fig:fig1}-\ref{fig:fig3} for up quarks ($q=u$) and in Figs.~\ref{fig:fig4}-\ref{fig:fig6} for down quarks ($q=d$). They are plotted as a function of $\bar x$ at different values of $t$ and $\xi$. 

In Figs.~\ref{fig:fig1} and \ref{fig:fig4} the $t$ dependence at $\xi=0$ is given for up and down quarks, respectively, for $H^q_T$, $E^q_T$, $\tilde H^q_T$. There is no $\tilde E^q_T$ because it vanishes identically being an odd function of $\xi$ as a consequence of time-reversal invariance~\cite{diehl01}.  $H^u_T$ and $\tilde H^u_T$ have opposite sign with respect to $H^d_T$ and $\tilde H^d_T$, with $H^u_T=-\oneq H^d_T$ at $t=0$ as expected from SU(6) symmetry. A comparison with results derived within the MIT bag model~\cite{jaffe92} is possible considering the forward limit $t=0$ of $H^q_T$, where $H^q_T$ reduces to the transversity $h^q_1$ (see below Eq.~(\ref{eq:forward})). The solid lines in Figs.~\ref{fig:fig1} and \ref{fig:fig4}, suitably scaled by the isospin factors $4/3$ and $-1/3$ for up and down quarks, respectively, almost overlap the result plotted in Fig. 2 of Ref.~\cite{jaffe92} for $h_1$ in the allowed region $0\le \bar x\le 1$. Indeed, the same weak $t$ dependence of $H^q_T$ as in the MIT bag model is found here. In contrast, the $t$ dependence affects the low-$\bar x$ region and is more pronounced in the cases of $E^q_T$ and $\tilde H^q_T$. For large $\bar x$ values the decay of all the distributions towards zero at the boundary $\bar x=1$ is almost independent of $t$. One can also notice that the combination $E^q_T + 2\tilde H^q_T$, more fundamental than $E^q_T$ itself when discussing spin densities in the transverse plane~\cite{diehl05}, is less sizable for down quarks than for up quarks also due to the oscillatory behavior of $E^d_T$.

The $\xi$ dependence for two values of $t\ne 0$ is given in Figs.~\ref{fig:fig2} and \ref{fig:fig3} for up quarks and in Figs.~\ref{fig:fig5} and \ref{fig:fig6} for down quarks. In all cases the GPDs vanish at $\bar x=\xi$ since in our approach they include the contribution of valence quarks only and we cannot populate the so-called ERBL region with $\vert\bar x\vert\le \xi$ where quark-antiquark pairs and gluons are important. Therefore, at low $\bar x$ this gives a strong $\xi$ dependence of the peak position of the distribution, but for large $\bar x$ the $\xi$ dependence turns out to be rather weak.

\section{The forward limit and the tensor charge}

In the forward limit $\Delta^\mu\rightarrow 0$ ($\bar x\to x$, with $x$ being the usual Bjorken variable), we immediately see from Eq.~(\ref{flip-amplitudes}) that only the quark GPDs $H_T^q$ can be measured. There they become equal to the quark transversity distributions $h_1^q(x)$. Although the quark GPDs $E_T^q$ and $\tilde H_T^q$ do 
not contribute to the scattering amplitude, they remain finite in the forward limit, whereas $\tilde E_T^q$ vanishes identically being an odd function of $\xi$ as already noticed~\cite{diehl01}.

As it is evident from Eqs.~(\ref{eq:gpd_oddampl}) and (\ref{eq:T3_ov}), the LCWF overlap representation of $H_T^q(x,0,0)$ for the valence-quark contribution is given by
\bea
\label{eq:forward}
H_T^q( x,0,0) & = & h^{q}_1(x)\nonumber\\
&=&\sum_{\mu^t_i\,\tau_i}\sum_{j=1}^3 \delta_{\tau_j\tau_q}\,
\mbox{sign}\,(\mu^{t}_j)
\int[dx]_3[d\vec{k}_\perp]_3\, \delta(x-x_j) 
\vert\Psi_{\uparrow}^{[f]}(\{x_i\},\{\vec{k}_{\perp,i}\};\mu^t_i,\tau_i\})
\vert^2.\nonumber\\
&&
\eea
This expression exhibits the well known probabilistic content of $h_1^q,$ being the probability to find a quark with spin polarized along the  transverse spin of a polarized nucleon minus the probability to find it 
polarized oppositely. Indeed $h^q_1$ is the counterpart in the transverse-polarization space
of the helicity parton distribution $g^q_1$ which measures the helicity asymmetry. As it was stressed by Jaffe and Ji~\cite{jaffe91}, in nonrelativistic situations where rotational and boost operations commute, 
one has $g^q_1=h^q_1$. Therefore the difference between $h^q_1$ and $g^q_1$  is a measure of the relativistic nature of the quarks inside the nucleon. In light-cone CQMs these relativistic effects are encoded in the Melosh rotations. With the help of Eqs.~(\ref{eq:a1})-(\ref{eq:a3}) and Eqs. (A.1), (A.3)-(A.4) of Ref.~\cite{BPT04}, we find
 \begin{eqnarray}
\label{eq:h1}
h^q_1(x)&=&\left(\frac{4}{3}\delta_{\tau_q 1/2}
-\frac{1}{3}\delta_{\tau_q -1/2}\right)
\int[dx]_3[d\vec{k}_\perp]_3\, \delta(x-x_3) 
\vert\tilde \psi_{\uparrow}(\{x_i\},\{\vec{k}_{\perp,i}\}\vert^2
\mathcal{M}_T,\\
\label{eq:g1}
g^q_1(x)&=&\left(\frac{4}{3}\delta_{\tau_q 1/2}
-\frac{1}{3}\delta_{\tau_q -1/2}\right)
\int[dx]_3[d\vec{k}_\perp]_3\, \delta(x-x_3) 
\vert\tilde \psi_{\uparrow}(\{x_i\},\{\vec{k}_{\perp,i}\}\vert^2
\mathcal{M},
\end{eqnarray}
where
\begin{eqnarray}
\label{eq:mt}
\mathcal{M}_T&=& \frac{(m+ x_3M_0)^2}
{(m+ x_3M_0)^2 + \vec{k}^2_{\perp,3}},
\\
& &\nonumber\\
\mathcal{M}&=& \frac{(m+ x_3M_0)^2 - \vec{k}^2_{\perp,3}}{(m+ x_3M_0)^2 + \vec{k}^2_{\perp,3}}.
\label{eq:m}
\end{eqnarray}
We note that in deriving the expression~(\ref{eq:mt}) for $\mathcal{M}_T$ we used the fact that the average squared momentum of the quarks in the $\hat x$ and $\hat y$ directions are the same. A similar result was already obtained in the relativistic CQM calculation of Ref.~\cite{CFT00}.

%%%%%%%%%%%%%%%%%%%%%%%%%%%%%  fig. 7  %%%%%%%%%%%%%%%%%%%%%%%%%%%%%%%%%%%%%%%
\begin{figure}[ht]
\begin{center}
\epsfig{file=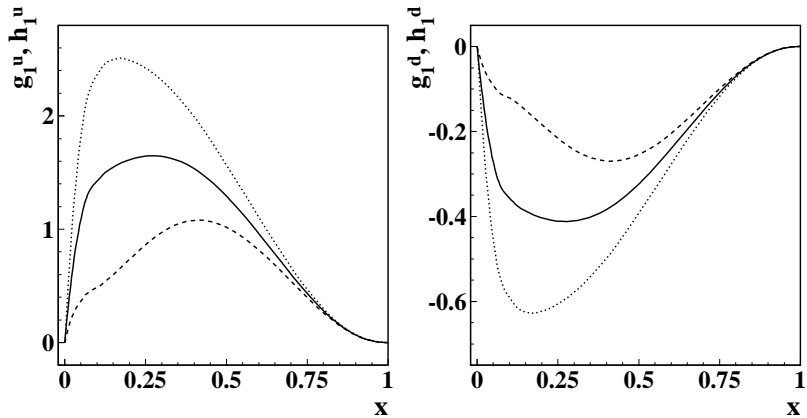,  width=26 pc}
\end{center}
\vspace{-0.4cm}
\caption{\small Helicity and transversity distributions for the $u$ (left panel) and $d$ (right panel) quark. The solid lines correspond to $h^q_1$, the dashed lines show $g^q_1$, and the dotted lines are the nonrelativistic results when Melosh rotations reduce to the identity ($h^q_1=g^q_1$).}
\label{fig:fig7}
\end{figure}

In Fig.~\ref{fig:fig7} the helicity and transversity distributions, $g_1$ and $h_1$, obtained as a forward limit of the corresponding GPDs calculated with the hypercentral CQM are compared and plotted together with the nonrelativistic result when Melosh rotations reduce to identity. The large difference between $g_1$ and $h_1$ shows how big is the effect of relativity. 

Recalling the expression for the unpolarized parton distribution $f^q_1$ obtained in Ref.~\cite{BPT03}, it is easy to see that the following relations hold
\begin{eqnarray}
\label{eq:soffer}
2h^u_1(x)=g^u_1(x)+\frac{2}{3}f^u_1(x),
& \quad&
2h^d_1(x)=g^d_1(x)-\frac{1}{3}f^d_1(x),
\end{eqnarray}
which are compatible with the Soffer inequality~\cite{soffer}. In the nonrelativistic limit, corresponding to $\vec{k}_\perp=0,$ and $\mathcal{M}_T=\mathcal{M}=1$, one obtains  $h^u_1=g^u_1=2/3 f^u_1$ and $h^d_1=g^d_1=-1/3 f^d_1.$ We note that the relations (\ref{eq:soffer}) generalize to the case of parton distributions the results obtained in Refs.~\cite{SS97,MaSS98} for the axial ($\Delta q$) and tensor ($\delta q$) charges, defined as
\begin{eqnarray}
\Delta q&=&\int_{-1}^1 d x \, g^q_1(x),\label{eq:axialcharge}\\
\delta q&=&\int_{-1}^1 d x \, h^q_1(x),\label{eq:tensorcharge}
\end{eqnarray}
respectively.
As a matter of fact, by calculating the first moment of the parton distributions in Eq.~(\ref{eq:soffer}), one recovers the following relations obtained in Refs.~\cite{SS97,MaSS98}
\begin{eqnarray}
2\delta q=\Delta q+\Delta q_{\rm{NR}}, \quad 
2\langle \mathcal{M}_T\rangle =\langle \mathcal{M}\rangle+1,
\end{eqnarray}
where $\Delta q_{\rm{NR}}$ is the axial charge in the nonrelativistic limit, i.e.
\begin{equation}
\Delta q_{\rm{NR}}=\left(\frac{4}{3}\delta_{\tau_q 1/2}
-\frac{1}{3}\delta_{\tau_q -1/2}\right).
\label{eq:tensor-nr}
\end{equation}

\begin{table}[h]
\caption{Valence contributions to the axial and tensor charge calculated within different SU(6)-symmetric quark models: the nonrelativistic quark model (NR), the harmonic oscillator model (HO) of Ref.~\cite{SS97}, and the hypercentral (HYP) model.}
\label{tab1}
\begin{center}
\begin{tabular}{crrrr}
\hline\hline
{} & $\, $ NR$\, $ &$\, $ HO$\, $ & $\, $HYP$\, $  \\ 
\hline 
$\Delta u\,$ & $\,4/3\,$  & $\,1.0\,$ & $\,0.61\,$ \\
$\Delta d\,$ & $\,-1/3\,$  & $\,-0.25\,$ & $\,-0.15\,$ \\
$\delta u\,$ & $\,4/3\,$     & $\,1.17\,$  & $\,0.97\,$ \\
$\delta d\,$ & $\,-1/3\,$ & $\,-0.29\,$ &$\,-0.24\,$ \\
\hline\hline  
\end{tabular}
\end{center}
\end{table}

%\bigskip

The nucleon tensor charge (\ref{eq:tensorcharge}) measures the net number of transversely polarized valence quarks in a transversely polarized nucleon~\cite{jaffe91,HeJi95}. Because of a nontrivial dynamical dependence of the rotation operators, it differs from the axial charge probed in high-energy processes and related to the net number of longitudinally polarized valence quarks in a longitudinally polarized nucleon. In the MIT bag model and demanding that $\Delta u - \Delta d = 1.257$, the tensor charge was fixed at $\delta u = 1.17$ and $\delta d= -0.29$ \cite{HeJi95}. These are remarkably the same numbers obtained in Ref.~\cite{SS97} with a simple harmonic oscillator wave function of the nucleon leading to its axial charge $g_A=1.25$. Furthermore, in contrast to the axial charge in the bag they are rather close to the nonrelativistic quark model result (see Eq.~(\ref{eq:tensor-nr}) and Table~\ref{tab1}) indicating less susceptibility to relativistic effects in the model. A detailed analysis of QCD sum rules in the presence of an external tensor field~\cite{HeJi96} gives $\delta u = 1.33\pm 0.53$ and $\delta d = 0.04\pm 0.02$ at the scale of the nucleon mass. This means that the up quarks dominate the contribution in a transversely polarized proton. In addition, the corresponding isovector ($g^v_T=\delta u - \delta d$) and isoscalar ($g^s_T=\delta u + \delta d$) tensor charges have similar size, $g^v_T =1.29\pm 0.51$ and $g^s_T=1.37\pm0.55$, and   $g^v_T$ is comparable in magnitude to the isovector axial charge $g_A$. The same trend of a dominating up-quark contribution is found also here in Table~\ref{tab1} with the LCWFs derived from the hypercentral CQM and can be understood by looking at our results in Figs.~\ref{fig:fig1} and \ref{fig:fig4}. However, the obtained numbers are closer to those derived in the nonrelativistic approach or the MIT bag than those predicted by QCD sum rules, with a nonnegligible negative contribution of the down quark. 
Although renormalization-scale dependent, the tensor charge is not affected by gluons. Therefore any discrepancy from what one could expect from QCD sum rules can be ascribed to the adopted LCWF that is here SU(6) symmetric.
%%%
When evolved in leading-order QCD from the intrinsic scale of the model ($Q_0^2=0.079$ GeV$^2$) to $Q^2=10$ GeV$^2$ the tensor charges become $\delta u =0.70$ and $\delta d=-0.17$ in the hypercentral CQM, within the range of values calculated in the different models considered in Ref.~\cite{barone} and in fair agreement with lattice QCD calculations~\cite{alikhan}.
%%%

Another quantity related to the forward limit of chiral-odd GPDS is the angular momentum $J^x$ carried by quarks with transverse polarization in the $\hat x$ direction in an unpolarized nucleon at rest. This quantity has recently been shown~\cite{Bur05} to be one half of the expectation value of the 
transversity asymmetry
\be
\langle \delta^x J^x_q\rangle = \langle J^x_{q,+\hat x} - J^x_{q,-\hat x}\rangle =
\oneh\left[ A_{T20} +2\tilde A_{T20} (0) + B_{T20}(0) \right],
\ee
where the invariant form factors $ A{T20},$ $\tilde A_{T20}$ and $B_{T20}$ are the second moments of the chiral-odd GPDs~\cite{diehl05,Bur05}:
\bea
A_{T20}(t) & = & \int_{-1}^1 dx\, x\, H_T(x,\xi,t), \nonumber\\
\tilde A_{T20}(t) & = & \int_{-1}^1 dx\, x\, \tilde H_T(x,\xi,t), \nonumber\\
B_{T20}(t) & = & \int_{-1}^1 dx\, x\, E_T(x,\xi,t) .
\eea
Using LCWFs derived from the hypercentral CQM we obtain 
\be
 \langle \delta ^x J^x_u\rangle=0.39, 
\quad \langle \delta ^x J^x_d\rangle=0.10,
\qquad\qquad \mbox{(HYP)}
\ee
while using the simple harmonic oscillator wave function of the nucleon as in Ref.~\cite{SS97}, we obtain much larger values:
\be
 \langle \delta ^x J^x_u\rangle=0.68, \quad \langle \delta ^x J^x_d\rangle=0.28.
\qquad\qquad \mbox{(HO)}
\ee
The same also occurs for the forward matrix element of $2\tilde H_T + E_T$, i.e.
\be
\kappa^q_T \equiv \int dx\left[2\tilde H^q_T(x,0,0) + E^q_T(x,0,0)\right].
\label{eq:kappat}
\ee
The quantity $\kappa^q_T$  describes how far and in which direction the average position of quarks with spin in the $\hat x$ direction is shifted in the $\hat y$ direction for an unpolarized nucleon~\cite{Bur05}. Thus $\kappa^q_T$ governs the transverse spin-flavor dipole moment in an unpolarized nucleon and plays a role similar to the anomalous magnetic moment $\kappa^q$ for the unpolarized quark distributions in a transversely polarized nucleon. As a matter of fact, we obtain
\be
\begin{array}{cclccl}
\kappa^u_T & = &  1.98, \quad \kappa^d_T & = & 1.17, \qquad\qquad & \mbox{(HYP)}\\
\kappa^u_T & = &  3.60, \quad \kappa^d_T & = & 2.36. \qquad\qquad & \mbox{(HO)}
\end{array}
\ee
Apart from their magnitude, the same sign of $\kappa^q_T$ is predicted in both models. This may have an impact on the Boer-Mulders function  $h_1^{\perp q}$ describing the asymmetry of the transverse momentum of quarks perpendicular to the quark spin in an unpolarized nucleon~\cite{BM98}. Since for $\kappa_T>0$ we expect that quarks polarized in the $\hat y$ direction should preferentially be deflected in the $\hat x$ direction, in accordance with the Trento convention~\cite{TN04} $\kappa^q_T>0$ would imply $h_1^{\perp q}<0$~\cite{Bur05}. Furthermore, keeping in mind that the magnitude of the quark anomalous magnetic moments $\kappa^q$ derived within the same approach are of the order of unity~\cite{BPT03}, the average Boer-Mulders function is predicted here larger than the average Sivers function $f_{1T}^{\perp q}\sim -\kappa^q$ describing the transverse momentum asymmetry of quarks in a transversely polarized target.
 
\section{Conclusions}

We have presented the general framework to calculate the overlap representation of chiral-odd generalized parton distributions using the Fock-state decomposition in the transverse-spin basis. The formalism has been applied to the case of light-cone wave functions obtained by considering only valence quarks in a constituent quark model. This limits the average longitudinal momentum fraction $\bar x$ to lie in the range between the skewness parameter $\xi$ and 1. The inclusion of quark-antiquark contributions is in principle possible following, e.g., the lines of Ref.~\cite{BPT05}.

For large $\bar x$ a weak dependence on $\xi$ and $t$ is found with opposite sign of $H^q_T$ and $\tilde H^q_T$ for up and down quarks. Different helicity and transversity distributions has been derived in the forward limit in agreement with the relativistic requirements and the Soffer inequality. A first estimate of the axial and tensor charges is also obtained confirming the different size and sign of the up and down quarks predicted within SU(6)-symmetric quark models. Furthermore, an analysis of the angular momentum carried by quarks with transverse polarization in an unpolarized nucleon leads to the prediction that the Boer-Mulders function describing the asymmetry of the transverse momentum of quarks perpendicular to the quark spin in an unpolarized nucleon could be larger than the average Sivers function describing the transverse momentum asymmetry of quarks in a transversely polarized target.

\bigskip

This research is part of the EU Integrated Infrastructure Initiative Hadronphysics Project under contract number RII3-CT-2004-506078 and was partially supported by the Italian MIUR through the PRIN Theoretical Physics of the Nucleus and the Many-Body Systems. 

\clearpage
%%%%%%%%%%%%%%%%%%%%%%%%%%%%%%%%%%%%%%%%%%%%%%%%%%%%%%%%%%%%%%%%%%%%%%%%%%%%
\begin{appendix}
\section{}
\label{appendix}
In this appendix we work out the summation over the spin and isospin variables appearing in the definition of the amplitudes. In the case of SU(6)-symmetric CQM wave functions, the summation over isospin variables gives
$\delta_{T_{12}0}\,\delta_{\tau_3 1/2} +
\delta_{T_{12}1}[\delta_{\tau_3 1/2} + 2\delta_{\tau_3-1/2}]/3$ for the
proton and 
$\delta_{T_{12}0}\,\delta_{\tau_3-1/2} +
\delta_{T_{12}1}[2\delta_{\tau_3 1/2} + \delta_{\tau_3 -1/2}]/3$ for the
neutron.
The summation over the spin variables is carried out in a similar way as in Ref.~\cite{BPT04} for the case of polarized GPDs, by using the explicit expressions of the Melosh-rotation matrices appearing in the initial and final light-cone wave function. As a result, one finds
\bea
& & T^q_{\lambda'\lambda } =
\frac{3}{2}\frac{1}{\sqrt{1-\xi^2}}\frac{1}{(16\pi^3)^2}
\int\prod_{1=1}^3 d\overline x_i\,\delta\left(1-\sum_{i=1}^3\overline x_i\right)
\,\delta(\overline x - \overline x_3) \nonumber\\
& & \qquad\times\int \prod_{i=1}^3
d^2\vec{k}_{\perp,i}\,\delta\left(\sum_{i=1}^3\vec{k}_{\perp,i}\right) \,
\tilde\psi^*(\{y'_i\},\{\vec{\kappa}'_{\perp,i}\})\,
\tilde\psi(\{y_i\},\{\vec{\kappa}_{\perp,i}\}) 
\nonumber\\
& & \qquad\times\delta_{\tau_q\tau_3}\left\{
X^{00}_{\lambda'\lambda}(\vec{\kappa}',\vec{\kappa})
\,\delta_{\tau_31/2} + \onet
X^{11}_{\lambda'\lambda}(\vec{\kappa}',\vec{\kappa})
[\delta_{\tau_31/2} + 2\delta_{\tau_3-1/2}]\right\},
\label{eq:x00_noflip}\label{eq:a1}\\ \nonumber
\eea
where
\be
\tilde\psi(\{y_i\},\{\vec{\kappa}_{\perp,i}\}) =
\left[\frac{1}{M_0}
\frac{\omega_1\omega_2\omega_3}{y_1 y_2 y_3}\right]
\psi(\vec{\kappa}_1,\vec{\kappa}_2,\vec{\kappa}_3).
\ee
For the functions $X$ in Eq.~(\ref{eq:x00_noflip}) we give only the expressions for the real part, since the contribution from the imaginary parts to $T^q_{\lambda'\lambda}$ is vanishing in the reference frame we are working with, where the momenta of the initial and final nucleon lie in the $x-z$ plane. As a result we have
\bea
\Re \Big (X^{00}_{++}(\vec{\kappa}',\vec{\kappa})\Big ) &=&
-
\Re \Big(X^{00}_{--}(\vec{\kappa}',\vec{\kappa}) \Big )\nonumber\\
&=&
\prod_{i=1}^3 N^{-1}(\vec{\kappa}'_i) N^{-1}(\vec{\kappa}_i)
 \Big[(A_1A_2 +\vec{B}_1\cdot\vec{B}_2)A_3\Big],
\label{eq:a3}\\ \nonumber
& &\\
\Re \Big(X^{11}_{++}(\vec{\kappa}',\vec{\kappa})\Big)& =&
-
\Re\Big(X^{11}_{--}(\vec{\kappa}',\vec{\kappa}) \Big)
=
\prod_{i=1}^3 N^{-1}(\vec{\kappa}'_i) N^{-1}(\vec{\kappa}_i)
\nonumber\\
& &
\times
\onet
\Big[ - (A_1A_2 +\vec{B}_1\cdot\vec{B}_2 -4B_{1,x}B_{2,x})A_3 
\nonumber\\
& &\quad+ 2 (A_1 B_{2,x}+A_2 B_{1,x}) B_{3,x}\nonumber\\
& &\quad + 2(B_{1,x} B_{2,z}+B_{1,z}B_{2,x})B_{3,y}\nonumber\\
& &\quad+2(B_{1,x} B_{2,y}+B_{1,y}B_{2,x})B_{3,z}\Big],
\label{eq:x11_noflip}
\\
\Re\Big(
X^{00}_{-+}(\vec{\kappa}',\vec{\kappa})\Big)
& = &
\Re \Big(X^{00}_{+-}(\vec{\kappa}',\vec{\kappa})\Big)
\nonumber\\
& = &
\prod_{i=1}^3 N^{-1}(\vec{\kappa}'_i) N^{-1}(\vec{\kappa}_i)
\Big[
(A_1A_2 +
\vec{B}_1\cdot\vec{B}_2)B_{3,y}\Big] ,\label{eq:a4}\\
\Re \Big(X^{11}_{-+}(\vec{\kappa}',\vec{\kappa})\Big)
& = &
\Re\Big(X^{11}_{+-}(\vec{\kappa}',\vec{\kappa})\Big)
= \prod_{i=1}^3 N^{-1}(\vec{\kappa}'_i) N^{-1}(\vec{\kappa}_i)
\nonumber\\
& &\times\onet
\Big[
(-A_1A_2 -\vec{B}_1\cdot\vec{B}_2 +4 B_{1,z}B_{2,z})B_{3,y}\nonumber\\
& &
\quad
+2(A_1 B_{2,z}+A_2 B_{1,z})B_{3,x}\nonumber\\
& &
\quad
+2(B_{1,x}B_{2,z}+B_{1,z}B_{2,x})A_3\nonumber\\
& &
\quad
+2(B_{1,y}B_{2,z}+B_{1,z}B_{2,y})B_{3,z}\Big],\\
\end{eqnarray}

In the above equations, $ N(\vec{\kappa})$,  $A_i$ and $\vec B_i,$ with $i=1,2,$  are defined as in Ref.~\cite{BPT03} and reported here for convenience
\begin{eqnarray}
N(\vec{\kappa}) &=& [(m+ yM_0)^2 + \vec{\kappa}^2_\perp]^{1/2}.
\\
A_i& =& (m+ y'_iM'_0)(m+ y_i  M_0) + \kappa'_{i,y} \kappa_{i,y} + \kappa'_{i,x} \kappa_{i,x},\qquad i=1,2
\\
B_{i,x}& = & - (m+ y'_iM'_0) \kappa_{i,y} + (m+ y_i  M_0) \kappa'_{i,y}, \qquad i=1,2,
\\
B_{i,y} &=& (m+ y'_iM'_0) \kappa_{i,x} -  (m+ y_i  M_0) \kappa'_{i,x},\qquad i=1,2,
\\
B_{i,z} &=& \kappa'_{i,x} \kappa_{i,y} - \kappa'_{i,y} \kappa_{i,x}, \qquad i=1,2,
,
\end{eqnarray}
while $A_3$ and $\vec{B}_3$ are given by
\begin{eqnarray}
A_3&=&  (m+ y'_3M'_0)(m+ y_3  M_0) +\kappa'_{3,y} \kappa_{3,y} -
 \kappa'_{3,x} \kappa_{3,x},
\\
B_{3,x} &=& (m+ y'_3M'_0) \kappa_{3,y} - (m+ y_3  M_0) \kappa'_{3,y}, 
\\
B_{3,y} &=& -(m+ y'_3M'_0) \kappa_{3,x} -  (m+ y_3  M_0) \kappa'_{3,x},
\\
B_{3,z} &=&  -\kappa'_{3,x} \kappa_{3,y} - \kappa'_{3,y} \kappa_{3,x}.
\end{eqnarray}
%with $y_i,\, y'_i,$ and $\kappa_i,\,\kappa'_i,$ for $i=1,2,3,$ defined 
%in Eqs.~(\ref{eq:finalstruck}) - (\ref{eq:initialspect}).

Analogously, the chiral-odd GPDs with flip of the transverse polarization of the active quark are obtained from the different matrix elements of the amplitude $\tilde T $ given explicitly by 
\bea
& & 
\tilde T^q_{\lambda'\lambda } =
\frac{3}{2}\frac{1}{\sqrt{1-\xi^2}}\frac{1}{(16\pi^3)^2}
\int\prod_{1=1}^3 d\overline x_i\,\delta\left(1-\sum_{i=1}^3\overline x_i\right)
\,\delta(\overline x - \overline x_3) \nonumber\\
& & \qquad\times\int \prod_{i=1}^3
d^2\vec{k}_{\perp,i}\,\delta\left(\sum_{i=1}^3\vec{k}_{\perp,i}\right) \,
\tilde\psi^*(\{y'_i\},\{\vec{\kappa}'_{\perp,i}\})\,
\tilde\psi(\{y_i\},\{\vec{\kappa}_{\perp,i}\}) 
\nonumber\\
& & \qquad\times\delta_{\tau_q\tau_3}\left\{
\tilde X^{00}_{\lambda'\lambda}(\vec{\kappa}',\vec{\kappa})
\,\delta_{\tau_31/2} + \onet
\tilde X^{11}_{\lambda'\lambda}(\vec{\kappa}',\vec{\kappa})
[\delta_{\tau_31/2} + 2\delta_{\tau_3-1/2}]\right\},
\label{eq:x00_tilde_noflip}\\ & &\nonumber
\eea
where
\bea
\Re\Big(\tilde X^{00}_{++}(\vec{\kappa}',\vec{\kappa})\Big) &=&
\Re\Big(\tilde X^{00}_{--}(\vec{\kappa}',\vec{\kappa}) \Big)
\nonumber\\
&=&
\prod_{i=1}^3 N^{-1}(\vec{\kappa}'_i) N^{-1}(\vec{\kappa}_i)
 \Big[(A_1A_2 +\vec{B}_1\cdot\vec{B}_2)\Big]\tilde A_3,
\\ \nonumber
& &\\
\Re\Big(\tilde X^{11}_{++}(\vec{\kappa}',\vec{\kappa})\Big)& =&
\Re\Big(X^{11}_{--}(\vec{\kappa}',\vec{\kappa}) \Big)
=
\prod_{i=1}^3 N^{-1}(\vec{\kappa}'_i) N^{-1}(\vec{\kappa}_i)
\nonumber\\
& &
\times
\onet
\Big[  (3 A_1A_2 -\vec{B}_1\cdot\vec{B}_2 )\tilde A_3 
\nonumber\\
& &\quad+ 2 (A_1 B_{2,x}+A_2 B_{1,x})\tilde B_{3,x}\nonumber\\
& &\quad + 2(A_1 B_{2,y}+A_2 B_{1,y})\tilde B_{3,y}\nonumber\\
& &\quad+2(A_1 B_{2,z}+A_2 B_{1,z})\tilde B_{3,z}\Big],
\label{eq:x11_tilde_noflip}
\\
\Re\Big(\tilde X^{00}_{-+}(\vec{\kappa}',\vec{\kappa})\Big)
& = &-
\Re \Big(\tilde X^{00}_{+-}(\vec{\kappa}',\vec{\kappa})\Big)
\nonumber\\
& = &
\prod_{i=1}^3 N^{-1}(\vec{\kappa}'_i) N^{-1}(\vec{\kappa}_i)
\Big[
(A_1A_2 +
\vec{B}_1\cdot\vec{B}_2)\tilde B_{3,y}\Big] ,\\
\Re \Big(\tilde X^{11}_{-+}(\vec{\kappa}',\vec{\kappa})\Big)
& = &-
\Re\Big(\tilde X^{11}_{+-}(\vec{\kappa}',\vec{\kappa})\Big)
= \prod_{i=1}^3 N^{-1}(\vec{\kappa}'_i) N^{-1}(\vec{\kappa}_i)
\nonumber\\
& &\times\onet
\Big[
(-A_1A_2 -\vec{B}_1\cdot\vec{B}_2 +4 B_{1,y}B_{2,y})\tilde B_{3,y}\nonumber\\
& &
\quad
+2(B_{1,x} B_{2,y}+B_{2,x} B_{1,y})\tilde B_{3,x}\nonumber\\
& &
\quad
+2(A_1B_{2,y}+A_2B_{1,y})\tilde A_3\nonumber\\
& &
\quad
+2(B_{1,y}B_{2,z}+B_{1,z}B_{2,y})\tilde B_{3,z}\Big],\\
\end{eqnarray}
where
\begin{eqnarray}
\tilde A_3&=&  \kappa'_{3,x}(m+ y_3  M_0) -\kappa_{3,x} (m+ y'_3M'_0),
\\
\tilde B_{3,x} &=&  -\kappa'_{3,x} \kappa_{3,y} - \kappa'_{3,y} \kappa_{3,x},
\\
\tilde B_{3,y} &=& (m+ y'_3M'_0) (m+ y_3  M_0)-\kappa'_{3,y}\kappa_{3,y}+ \kappa'_{3,x} \kappa_{3,x},
\\
\tilde B_{3,z} &=&  -(m+ y'_3M'_0)\kappa_{3,y} -(m+ y_3  M_0)
\kappa'_{3,y}.
\end{eqnarray}
\end{appendix}
%%%%%%%%%%%%%%%%%%%%%%%%%%%%%  the bibliography  %%%%%%%%%%%%%%%%%%%%%%%%%%%%

\clearpage
%%%%%%%%%%%%%%%%%%%%%%%%%%%%%

\begin{thebibliography}{99}

\bibitem{barone}
V.~Barone, A.~Drago, P.G.~Ratcliffe, Phys. Rep. 359, 1 (2002); V.~Barone, P.G.~Ratcliffe, {\it Transverse spin physics\/}, World Scientific, Singapore, 2003.

\bibitem{varia}
W.~Vogelsang, hep-ph/0309295; A.~Metz, hep-ph/0412156; V.~Barone, hep-ph/0502108.

\bibitem{ralston}
J.~Ralston, D.~Soper, Nucl. Phys. B 152, 109 (1979).

\bibitem{jaffe91}
R.L.~Jaffe, X.~Ji, Phys. Rev. Lett. 67, 552 (1991).

\bibitem{ralston2}
L.~Cortes, B.~Pire, and J.P.~Ralston, Z. Phys. C 55, 409 (1992).

\bibitem{jaffe92}
R.L.~Jaffe, X.~Ji, Nucl. Phys. B 375, 527 (1992).


\bibitem{pax}
P.~Lenisa, F.~Rathmann, for the PAX collaboration, hep-ex/0505054.

\bibitem{artru}
X.~Artru, M.~Mekhfi, Z. Phys. C 45, 669 (1990).

\bibitem{collins}
J.C.~Collins, Nucl. Phys. B 396, 161 (1993).

\bibitem{tang}
R.L.~Jaffe, X.~Jin, J.~Tang, Phys. Rev. Lett. 80, 1166 (1998).

\bibitem{radici}
M. Radici, R. Jakob, A. Bianconi, Phys. Rev. D 65, 074031 (2002).

\bibitem{bacchetta}
A.~Bacchetta, M.~Radici, Phys. Rev. D 67, 094002 (2003).

\bibitem{muller}
D. M\"uller, D. Robaschik, B. Geyer, F.M. Dittes, J. Ho\u rej\u si, Fortsch. Phys. 42, 101 (1994).

\bibitem{radyushkin96}
A.V. Radyushkin, Phys. Lett. B 380 (1996) 417; Phys. Lett. B 385, 333 (1996).

\bibitem{ji78}
Xiangdong Ji, Phys. Rev. Lett. 78, 610 (1997); J. Phys. G 24, 1181 (1998).

\bibitem{goeke}
K.~Goeke, M.V.~Polyakov, M.~Vanderhaeghen, Prog. Part. Nucl. Phys. 47, 401 (2001). 

\bibitem{diehl03}
M. Diehl, Phys. Rep. 388, 41 (2003).

\bibitem{ji04}
Xiangdong Ji, Ann. Rev. Nucl. Part. Sci. 54, 413 (2004).

\bibitem{BR05}
A.V.~Belitsky, A.V.~Radyushkin, hep-ph/0504030.

\bibitem{HJ98}
P. Hoodbhoy, X. Ji, Phys. Rev. D 58, 054006 (1998).

\bibitem{diehl01}
M. Diehl, Eur. Phys. J. C 19, 485 (2001).

\bibitem{exp_pl1}
Jefferson Lab. experiment in Hall A,  E03-106;
Jefferson Lab. experiment in Hall A, E00-110;
Jefferson Lab. experiment in Hall B, E01-113;

\bibitem{exp_pl2}
N. D'Hose {\it et al.}, (COMPASS Coll.), Eur. Phys. J. A19, s01, 47 (2004).

\bibitem{exp1}
HERMES Collaboration, A. Airapetian {\it et al.}, Phys. Rev. Lett. 87, 182001 
(2001).
\bibitem{exp2}
CLAS Collaboration, S. Stepanyan {\it et al.}, Phys. Rev. Lett. 87, 182002 (2001);
CLAS Collaboration, C.Hadjidakis {\it et al.}, Phys. Lett. B 605, 256 (2005).

\bibitem{ivanov}
D.Yu.~Ivanov, B.~Pire, L.~Szymanowski, O.V.~Teryaev, Phys. Lett. B 550, 65 (2002); Phys. Part. Nucl. 35, 67 (2004).

\bibitem{diehl05}
M.~Diehl, P.~H\"agler, hep-ph/0504175.

\bibitem{Bur05}
M. Burkardt, hep-ph/0505189.

\bibitem{gockeler}
M. G\"ockeler {\it et al.}  (QCDSF Collaboration), Nucl. Phys. A 755, 537c (2005).

\bibitem{ADH97}
S. Aoki, M. Doui, T. Hatsuda, and Y. Kuramashi, Phys. Rev. D 56, 433 (1997).

\bibitem{scopetta}
S.~Scopetta, hep-ph/0509287.

\bibitem{diehl2}
M.~Diehl, Th.~Feldmann, R.~Jakob, P.~Kroll, Nucl. Phys. B 596, 33 (2001).

\bibitem{brodsky}
S.J.~Brodsky, M.~Diehl, D.S.~Hwang, Nucl. Phys. B 596, 99 (2001).

\bibitem{BPT03}
 S. Boffi, B. Pasquini and M. Traini, Nucl. Phys. B 649, 243 (2003).

\bibitem{BPT04}
S. Boffi, B. Pasquini and M. Traini, Nucl. Phys. B 680, 147 (2004).

\bibitem{BHMS}
S.J. Brodsky, D.S.~Hwang, B.-Q. Ma,  I. Schmidt, Nucl. Phys. B 593, 311 (2001).

\bibitem{burkardt05}
M.~Burkardt, hep-ph/0510249.

\bibitem{brodsky2002}
S.J. Brodsky, P. Hoyer, N. Marchal, S. Peign\'e, F. Sannino, Phys. Rev. D 65, 114025 (2002).

\bibitem{belitsky2003}
A.V. Belitsky, X. Ji, F. Yuan, Nucl. Phys. B 656, 165 (2003).

\bibitem{Jaffe}
R.~L.~Jaffe,
%``Spin, twist and hadron structure in deep inelastic processes,''
hep-ph/9602236.

\bibitem{FTV99}
P. Faccioli, M. Traini and V. Vento, Nucl. Phys. A 656, 400 (1999).

\bibitem{CFT00}
F. Cano, P. Faccioli, M. Traini, Phys. Rev. D 62, 094018  (2000) .

\bibitem{soffer}
J. Soffer, Phys. Rev. Lett. 74, 1292 (1995).

\bibitem{SS97}
I. Schmidt, J. Soffer, Phys. Lett. B 407, 331 (1997).

\bibitem{MaSS98}
B.-Q. Ma,  I. Schmidt, J. Soffer, Phys. Lett. B 441, 461 (1998).

\bibitem{HeJi95}
H. He and X. Ji, Phys. Rev. D 52, 2960 (1995).

\bibitem{HeJi96}
H. He and X. Ji, Phys. Rev. D 54, 6897 (1996).

\bibitem{alikhan}
A.~Ali Khan {\it et al.}  (QCDSF-UKQCD Collaboration), hep-lat/0409161.

\bibitem{BM98}
D.~Boer, P.J.~Mulders, Phys. Rev. D 57, 5780 (1998).

\bibitem{TN04}
A.~Bacchetta, U.~D'Alesio, M.~Diehl, C.A.~Miller, Phys. Rev D 70, 117504 (2004).

\bibitem{BPT05}
B. Pasquini, M. Traini, and S. Boffi, Phys. Rev. D 71, 034022 (2005).

%%%%%%%%%%%%%%%%%%%%%%%%%%%%%
\end{thebibliography}
\end{document}